\documentclass[sigconf]{acmart}

\usepackage{multirow}
\usepackage[T1]{fontenc}

\AtBeginDocument{%
  }

\setcopyright{cc}
\setcctype{by}
\copyrightyear{2026}
\acmYear{2026}
\acmDOI{10.1145/3799682.3840101}
\acmConference[CIKM '26]{Proceedings of the 35th ACM International Conference on Information and Knowledge Management}{November 07--11, 2026}{Rome, Italy}
\acmBooktitle{Proceedings of the 35th ACM International Conference on Information and Knowledge Management (CIKM '26), November 07--11, 2026, Rome, Italy}
\acmISBN{979-8-4007-2539-5/2026/11}

\begin{document}

\title{SSR-GRPO: Integrating Supervision and Semantic IDs into Reinforcement Learning for Dense Retrieval in E-commerce}

\author{Guangxin Song}
\email{1220635451@qq.com}
\affiliation{%
  \institution{Alibaba Group}
  \city{Hangzhou}
  \country{China}
}

\author{Xing Fang}
\email{fangxing.fx@taobao.com}
\affiliation{%
  \institution{Alibaba Group}
  \city{Hangzhou}
  \country{China}
}

\author{Mingmin Jin}
\email{jimmy.jmm@taobao.com}
\affiliation{%
  \institution{Alibaba Group}
  \city{Hangzhou}
  \country{China}
}

\author{Jing Wang}
\email{jing.wangj1@taobao.com}
\affiliation{%
  \institution{Alibaba Group}
  \city{Hangzhou}
  \country{China}
}

\author{Bokang Wang}
\email{wangbokang.wbk@alibaba-inc.com}
\affiliation{%
  \institution{Alibaba Group}
  \city{Hangzhou}
  \country{China}
}

\author{Zhentao Song}
\email{zhentao.szt@alibaba-inc.com}
\affiliation{%
  \institution{Alibaba Group}
  \city{Hangzhou}
  \country{China}
}

\author{Junjie Bai}
\email{baijunjie.bjj@taobao.com}
\affiliation{%
  \institution{Alibaba Group}
  \city{Hangzhou}
  \country{China}
}

\author{Jianbo Zhu}
\email{zhujianbo.zjb@taobao.com}
\affiliation{%
  \institution{Alibaba Group}
  \city{Hangzhou}
  \country{China}
}

\renewcommand{\shortauthors}{Song et al.}

\begin{abstract}
Embedding-based retrieval (EBR) is a prevailing method in modern e-commerce search, retrieving relevant items from billions of candidates in response to users' queries. While recent methods often fine-tune large language models (LLMs) for representation learning, they typically lack robust mechanisms for handling complex and implicit semantics.

To address this, Retrieval-GRPO (R-GRPO) was recently proposed as a multi-objective Reinforcement Learning (RL) framework for dense retrieval. It computes rewards using top-$K$ items dynamically retrieved per query. While this approach partially mitigates retrieval challenges, the limited candidate pool sampled per batch often introduces noisy samples into the top-$K$ results. Moreover, the LLM used for computing relevance rewards in R-GRPO is trained with a RL framework similar to R-GRPO itself. This inherent similarity makes it difficult to provide unbiased and impartial assessment results. To tackle these issues, we propose Supervised Retrieval-GRPO with Semantic Identifiers (SSR-GRPO). Specifically, our method first proposes a dual-perspective framework for relevance assessment. It leverages both Semantic Identifiers (SIDs) produced by quantization learning and dense representation vectors to generate more unbiased relevance scores. Furthermore, leveraging the hierarchical similarity relationships of the generated SIDs, we mine a set of hard negative samples that serve two purposes: (1) to design a masking function integrated into R-GRPO, effectively filtering intra-group noisy samples; and (2) to construct a Retrieval-DPO task composed of positive and negative sample pairs, enabling the model to capture fine-grained semantic distinctions from a pair-wise perspective. By integrating these optimization strategies, we propose SSR-GRPO. Extensive offline and online experiments validate SSR-GRPO's effectiveness, and it has been deployed on a large-scale e-commerce platform.
\end{abstract}




\maketitle

\section{Introduction}
\label{sec:intro}

Dense retrieval, also called embedding-based retrieval (EBR), is fundamental to modern e-commerce search \cite{sondhi2018taxonomy}. It encodes queries and items into a shared embedding space \cite{huang2013learning,zhang2020towards,kong2022multi} where semantic similarity is computed via distance metrics like cosine similarity. While early approaches used encoder-only architectures like BERT and RoBERTa \cite{devlin2019bert,liu2019roberta}, recent LLM-based systems (e.g., Gemini \cite{lee2025gemini}, Qwen3 \cite{zhang2025qwen3}) employ techniques like hard negative mining \cite{robinson2020contrastive} and LLM-generated data synthesis \cite{wu2024llm}. However, most methods still follow BERT-era training paradigms.

Recent work explores integrating Reinforcement Learning (RL) with LLMs for retrieval tasks. Jiang et al. \cite{jiang2025deepretrieval} proposed using RL to train LLMs for query generation/rewriting, achieving strong retrieval performance. This reflects a broader trend of applying policy optimization to information retrieval. In e-commerce dense retrieval, Retrieval-GRPO (R-GRPO) \cite{liu2025taosearchemb} leverages GRPO to train semantic retrieval models, capturing complex semantics through multi-objective rewards while eliminating manual hard negative mining. Specifically, at each training step, R-GRPO retrieves the top-K most similar products from the current batch as hard negative samples.
However, the products within each training batch represent only a limited subset of the overall product pool. Thus, there is no guarantee that the top-K items retrieved at each step are all valuable. Moreover, the retrieved top-K set may contain easy negative samples, which can introduce noise and misguidance into the model training process. Similar issues exist in the domain of LLMs. To address this, DeepSeekV3.2 proposed Off-policy Sequence Masking \cite{liu2025deepseek} strategy to mitigate the problem. It stabilizes training by masking sequences causing policy divergence. Beyond the noise of top-K items, R-GRPO's reward mechanism introduces another significant concern: the calculation of relevance reward relies on TaoSR1 \cite{dong2025taosr1}, a 42B-MoE LLM. This dependency presents two main drawbacks: (1) The introduction of a high-parameter model for inference requires substantial extra computational resources during training; (2) Since TaoSR1 is trained using a reinforcement learning framework closely mirroring the one used in R-GRPO for query-item semantic matching, it tends to produce correlated relevance judgments. This correlation makes it difficult to provide unbiased and highly discriminative assessments.

To tackle these challenges, we propose a dense-sparse dual-perspective relevance reward evaluation method to replace the single LLM inference assessment. In the dense component, we utilize representation embeddings from a multimodal relevance LLM to compute inner product scores. In the sparse perspective, we employ Semantic Identifiers (SIDs) generated via Residual Quantized Variational Autoencoder (RQ-VAE) for items and through a next-token-prediction task for queries. The relevance score is then calculated based on the matching depth between the query and item SIDs. Furthermore, we design a joint training framework that enables the model to learn fine-grained semantic information and implicit patterns from multiple perspectives. Specifically, we introduce the Retrieval-DPO (R-DPO) approach to enable the model to learn representation differences from a pair-wise perspective. This is achieved by mining hard negative samples through the hierarchical similarity relationships of SIDs, combined with traditional hard negative mining strategies to construct a novel R-DPO loss and a masking function that provides explicit supervision signals to R-GRPO.

To the best of our knowledge, we are the first to leverage SIDs generated by RQ-VAE for mining hard negative samples in dense retrieval, and the first to integrate SID-based matching into reinforcement learning reward calculation. In summary, our proposed SSR-GRPO introduces the following innovations:
\begin{itemize}
    \item We propose a dual-perspective relevance reward calculation method to enhance the reliability of relevance assessment.
    \item We introduce a hard negative sample mining strategy based on SID similarity. The newly mined samples are utilized to establish a joint training paradigm integrating R-DPO and R-GRPO, thereby enhancing the generalization capability of the reinforcement learning framework.
    \item We design a hard negative-based masking function for R-GRPO to reduce noise from simple negatives.
    \item By integrating the aforementioned components, we develop SSR-GRPO, a novel semantic retrieval paradigm rigorously validated through extensive offline and online A/B testing.
\end{itemize}
\section{Related Work}
\label{sec:related}

\subsection{Dense Retrieval}
Embedding-based retrieval (EBR) based on dual-encoder architectures has been fundamental to search systems, enabling efficient similarity computation between queries and items \cite{huang2020embedding,liu2021que2search}. Early work focused on personalized representations \cite{li2021embedding} and BERT-based models \cite{guo2020detext,niu2020dual}, while SimCSE \cite{gao2021simcse} introduced contrastive learning for dense retrieval. Recently, LLMs have emerged as powerful backbones for generating universal semantic representations, with models like RepLLaMA \cite{ma2024fine} and newer state-of-the-art embeddings \cite{lee2024nv,zhang2025qwen3} demonstrating strong retrieval capabilities.

\subsection{Reinforcement Learning in Information Retrieval}
Reinforcement learning has gained traction for improving LLM performance and controllability. While PPO \cite{schulman2017proximal} dominated early RLHF approaches, recent methods like DPO \cite{rafailov2023direct} and GRPO \cite{shao2024deepseekmath} offer more efficient alternatives. In retrieval, RL has been applied to query generation \cite{jiang2025deepretrieval}, relevance reasoning \cite{zeng2025optimizing}, and dense retrieval optimization \cite{liu2025taosearchemb}. DeepRetrieval \cite{jiang2025deepretrieval} combines GRPO with LLM to enable unsupervised query generation, offering a more efficient training paradigm for information retrieval. Zeng et al. \cite{zeng2025optimizing} formulate relevance modeling as a reasoning task and introduce a RL-based training framework to enhance the grounded reasoning capabilities of generative relevance models. Retrieval-GRPO \cite{liu2025taosearchemb} applies GRPO strategy to dense retrieval, which eliminates the dependency on labor-intensive hard negative sample mining while mitigating the seesaw effect inherent in multi-objective optimization.

\subsection{Semantic IDs in LLMs and Information Retrieval}

Generative Retrieval \cite{rajput2023recommender,liu2025generative} represents products using Semantic Identifiers (SIDs)—discrete token sequences that encode product semantics into generatable identifiers \cite{wu2024hi,lin2025order}. SIDs are typically generated through an encoder-quantization framework \cite{esser2021taming,lee2022autoregressive} that maps multimodal features into discrete spaces. This approach enables semantically similar products to receive similar SIDs, facilitating more accurate matching in LLM-based e-commerce retrieval systems.
 \section{Methods}
\label{sec:method}
\begin{figure*}[htbp]
    \centering
    \includegraphics[width=0.95\linewidth]{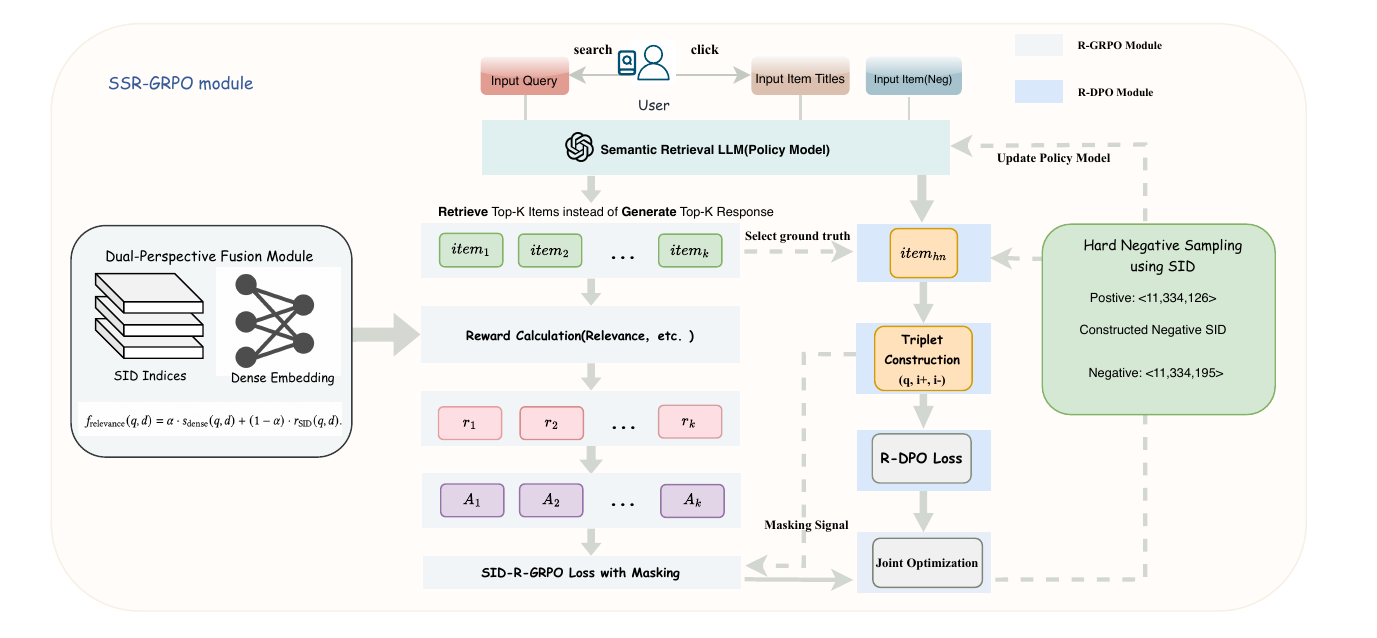} 
    \caption{The SSR-GRPO overall training framework consists of four components: (1) a semantic retrieval LLM trained via SFT; (2) R-GRPO module containing the dual-perspective reward calculation based on SIDs and masking mechanism; (3) R-DPO Optimization using hard negatives derived from hierarchical SIDs; (4) SID-R-GRPO and R-DPO joint training based on dynamic weighting.}
    \Description{The SSR-GRPO overall training framework }
    \label{fig:ssr}
\end{figure*}

\subsection{Preliminary}
Dense retrieval aims to map queries and items into a shared embedding space for efficient relevance matching via geometric proximity. Given query $q_i$ and item collection $\mathcal{D}$, we employ a dual-encoder model $f_\theta(\cdot)$ to encode them as:
\begin{equation}
\mathbf{q_i} = f_\theta(q_i), \quad \mathbf{d_j} = f_\theta(d_j) \in \mathbb{R}^{D}.
\end{equation}
Both query and item embeddings are \text{L2-normalized} to unit vectors, i.e., $\|\mathbf{q_i}\|_2 = 1$ and $\|\mathbf{d_j}\|_2 = 1$. Under this normalization, retrieval computes similarity scores using inner product:
\begin{equation}
s(q_i, d_j) = \langle \mathbf{q_i}, \mathbf{d_j} \rangle,
\end{equation}
which is equivalent to computing the cosine similarity between the two vectors. The system then retrieves top-K relevant items:
\begin{equation}
\mathcal{D}_{q_i} = \text{TopK}_{d_j \in \mathcal{D}} s(q_i,d_j).
\end{equation}

\subsection{SFT}
\label{sec:infonce-loss}
Our framework employs a two-stage training process: supervised fine-tuning (SFT) followed by SSR-GRPO. In SFT, $(q,d)$ positive pairs and in-batch negatives \cite{liu2021que2search, li2021embedding} are used to pull query representations close to relevant items while pushing them away from negative items via InfoNCE loss \cite{oord2018representation}:

\begin{equation}
\mathcal{L}_{\text{InfoNCE}} = -\log \frac{\exp(s(q_i, d_j^+)/\tau)}{\sum_{d_j \in \mathcal{B}} \exp(s(q_i, d_j)/\tau)},
\label{eq:infonceloss}
\end{equation}
where $s(\cdot)$ is the similarity score, $\mathcal{B}$ the in-batch negative samples, and $\tau$ a temperature parameter \cite{li2023learning,wang2023learning} smoothing the fitted data distribution.

\subsection{SSR-GRPO}
\label{sec:supervised-sid-r-grpo}

\subsubsection{Reward Function in R-GRPO}
\label{sec:sid-retrieval-grpo}
The reward function in SSR-GRPO follows R-GRPO's multi-objective formulation, combining relevance, quality, and exclusivity components as in Eq.~\eqref{eq:reward_components}:
\begin{equation}
    r = f_{\text{relevance}}(q, d) + g_{\text{quality}}(d) + h_{\text{exclusivity}}(q, d)
\label{eq:reward_components}.
\end{equation}
While $g_{\text{quality}}$ and $h_{\text{exclusivity}}$ remain identical to R-GRPO, we propose a novel dual-perspective relevance assessment for $f_{\text{relevance}}$. This integrates: (1) a dense signal via inner product between multi-modal embeddings $s(\mathbf{e}_q, \mathbf{e}_d)$, and (2) a sparse signal using SIDs, where relevance is scored based on hierarchical matching depth. The SID construction and scoring strategy are detailed next.

\subsubsection{Item SID Learning via Contrastive RQ-VAE.}
\label{subsec:item-sid-rqvae}
We adopt a query-bridged contrastive quantization framework to learn hierarchical Semantic Identifiers (SIDs) for items. The process begins with pre-trained relevance embeddings $\mathbf{e}_q = \text{RelevanceEmb}_q(q)$ and $\mathbf{e}_d = \text{RelevanceEmb}_d(d)$, obtained from a dual-tower model trained on large-scale interaction data. Lightweight MLP encoders then project these embeddings into a shared latent space: $\mathbf{z}_q = \text{Enc}_q(\mathbf{e}_q)$, $\mathbf{z}_d = \text{Enc}_d(\mathbf{e}_d)$.
A shared Residual Quantized Variational Autoencoder (RQ-VAE) with $L$(e.g., $L$=3) levels of codebooks $\{C_1, C_2, \ldots, C_L\}$ performs residual quantization on the item representations:
\begin{equation}
\mathbf{r}^0 = \mathbf{z}_d, \quad c^\ell = \arg\min_{e \in C_\ell} \|\mathbf{r}^{\ell-1} - e\|_2^2, \quad \mathbf{r}^\ell = \mathbf{r}^{\ell-1} - c^\ell,
\end{equation}
producing the item SID as $\text{SID}_i = [k_1, k_2, \ldots, k_L]$. The RQ-VAE is optimized through a joint training objective that combines a commitment loss, a reconstruction loss and an InfoNCE contrastive loss :
\begin{equation}
\mathcal{L} = \lambda_{\text{infoNCE}} \mathcal{L}_{\text{InfoNCE}} + \lambda_{\text{commit}} \mathcal{L}_{\text{commit}} + \lambda_{\text{recon}} \mathcal{L}_{\text{recon}}.
\end{equation}
Crucially, the InfoNCE loss encourages items co-retrieved by the same query to share consistent SIDs, turning geometric clustering into relevance-aware hierarchical partitioning.

\subsubsection{Query SID Generation via Generative LLM.}
We formulate the generation of Query SIDs as a next-token-prediction (NTP) task, fine-tuned on an auto-regressive Large Language Model (LLM). The training data is constructed from historical search logs, where each query is paired with the Semantic Identifiers (SIDs) of items that received positive user interactions—such as clicks, add-to-cart actions, and purchases. These interacted item SIDs serve as the ground-truth targets for the model. The model is trained to minimize the negative log-likelihood loss:
\begin{equation}
\mathcal{L}_{\text{SFT}} = -\sum_{t=1}^{L} \log P(k_t \mid q, k_{<t}; \theta),
\end{equation}
where $k_t$ denotes the codebook index at level $t$ and $\theta$ represents the LLM parameters. During inference, beam search is employed to generate the top-$K$ most probable SID sequences $\{\text{SID}_q^1, \ldots, \text{SID}_q^K\}$. This approach naturally captures the one-to-many mapping from queries to relevant item clusters, effectively addressing query ambiguity.

\subsubsection{Dual-Perspective Relevance Reward with SID Fusion}
\label{subsec:sid-scoring}

Given query SIDs $\{\text{SID}_q^j\}_{j=1}^K$ and item SID $\text{SID}_i=[k_1^i, k_2^i, k_3^i]$, we compute the maximum matching depth:
\begin{equation}
\ell^*(q, d) = \max_{j=1}^{K} \max \left\{\ell \in \{1, \ldots, L\} : k_t^{q,j} = k_t^i, \; \forall t \leq \ell\right\},
\end{equation}
where a deeper shared prefix signifies finer-grained semantic coherence. This discrete hierarchical score is then mapped to a scalar value:
\begin{equation}
r_{\text{SID}}(q, d) =
\begin{cases}
0.0 & \text{if } \ell^* = 0 \quad \text{(No match)}, \\
0.25 & \text{if } \ell^* = 1 \quad \text{(Coarse-level match)}, \\
0.5 & \text{if } \ell^* = 2 \quad \text{(Mid-level match)}, \\
1.0 & \text{if } \ell^* = 3 \quad \text{(Fine-level match)}.
\end{cases}
\end{equation}
The final relevance reward is obtained by fusing the dense embedding similarity with the sparse SID matching score, forming the core of our dual-perspective assessment:
\begin{equation}
\label{eq:relevance_fusion}
f_{\text{relevance}}(q, d) = \alpha \cdot r_{\text{SID}}(q, d) +  (1 - \alpha) \cdot s_{\text{dense}}(q, d).
\end{equation}
Here, $\alpha \in [0, 1]$ is a balancing hyperparameter.
This dual-perspective mechanism leverages the complementary strengths of both representations: the dense embeddings provide a smooth, continuous measure of overall semantic similarity, while the discrete SIDs offer sharp, hierarchical discrimination based on key attribute alignments. Together, they form a more resilient and accurate relevance reward, crucial for guiding the policy optimization in R-GRPO.

\subsubsection{Hard Negative Mining and Masking Mechanism}
\label{sec:supervised-hn-masking}

We introduce a novel method which leverages SIDs introduced in Section~\ref{subsec:item-sid-rqvae} for hard negative mining, combined with a masking mechanism. This approach provides explicit supervision signals to augment and stabilize the R-GRPO process.
As for the hard negative mining, we select hard negatives from two complementary perspectives to address different aspects of retrieval challenges:

\textbf{Type I: Commercial Performance-based Hard Negatives.} 
We target items that consistently reach the pre-ranking stage but fail final exposure due to low scores from subsequent ranking stages. These ranking models employ sophisticated cross-features and sequential modeling, providing accurate conversion efficiency assessments. Thus, low-scoring items typically exhibit genuinely poor conversion performance (e.g., low CTR/CVR). Although semantically relevant to queries, they demonstrate weak commercial performance. Using them as hard negatives helps the model better discriminate between semantically relevant but commercially weak items and those with strong conversion potential.

\textbf{Type II: Semantic Fine-grained Hard Negatives.}
To capture subtle semantic distinctions, we leverage hierarchical SIDs. For a positive item with $\text{SID}_{i^+} = [k_1, k_2, k_3]$, we select negatives sharing the first two SID levels ($k_1, k_2$) but differing at the third ($k_3$). This creates challenging negatives that are semantically proximate at coarse granularity yet distinct in fine-grained attributes, forcing the model to learn nuanced semantic boundaries essential for e-commerce relevance.

To provide explicit supervision signals for filtering out trivial negative samples within the top-K items in R-GRPO, we design a novel masking mechanism using the mined hard negatives as quality benchmarks. In the GRPO framework, these hard negatives (denoted as $d_{\text{hn}}$) serve as quality benchmarks. We apply a masking function designed to filter out easy negative samples by removing any top-K item whose relevance reward score falls below the benchmark. The masking function is defined as:
\begin{equation}
M_{i} = \begin{cases}
    0, & f_{\text{relevance}}(q,d_{i}) < f_{\text{relevance}}(q,d_{\text{hn}}) \\
    1, & \text{otherwise},
\end{cases}
\label{eq:ssr_mask_definition}
\end{equation}
where $M_i$ is a binary function, if the current item's relevance reward is lower than the relevance reward of the hard negative, then $M_i=0$ (masked out), otherwise $M_i=1$.

\subsubsection{SID-R-GRPO}
\label{sec:sid-grpo}
Combining the dual-perspective relevance reward (Section~\ref{subsec:sid-scoring}) with the masking mechanism (Section~\ref{sec:supervised-hn-masking}), we obtain SID-R-GRPO. In SID-R-GRPO, we treat the top-$K$ items retrieved by the semantic retrieval LLM illustrated in Figure \ref{fig:ssr} as candidates. Using the reward models integrating SIDs from Section~\ref{sec:sid-retrieval-grpo}, we compute real-time rewards and penalize low-scoring items to dynamically adjust predictions. Candidates are gathered across devices into a large batch $\hat{B}$ ($k \ll \hat{B}$), with top-$k$ selected for reward computation. The inter-group advantage $A_i$ among top-$K$ items under the same query reflects multi-perspective user preferences, determining reward/penalty assignments. The SID-R-GRPO loss is defined as:

\begin{equation}
\begin{aligned}
\mathcal{L}_{\text{SID-R-GRPO}} & = -\frac{1}{K} \sum_{i=1}^{K}\Bigg\{ 
 \min\Bigg(\pi_\theta(s(q,d_i)|q,d_i)\hat{A}_{i}, \\
& \text{clip}\left(\pi_\theta(s(q,d_i)|q,d_i), 1-\varepsilon, 1+\varepsilon\right)\hat{A}_{i}\Bigg)M_{i}\Bigg\} + \beta \mathcal{L}_{\text{InfoNCE}},
\end{aligned}
\label{sidrgrpo_loss}
\end{equation}
where
\begin{equation}
M_{i} = \begin{cases}
    0, & f_{\text{relevance}}(q,d_{i}) < f_{\text{relevance}}(q,d_{\text{hn}}) \\
    1, & \text{otherwise},
\end{cases}
\label{ssrgrpo_masking}
\end{equation}

\begin{equation}
    A_{i} = \frac{r_{i} - mean(\mathbf{r})}{std(\mathbf{r})},
\end{equation}
and $K$ is the size of the sample set composed of top-$k$ for the current query, $\pi_\theta$ is the distribution of the policy model, $A_i$ represents the advantage value of the $i$-th item in top-$k$ computed by multi-objective rewards. The collection $\mathbf{r} = \{r_1, r_2, \ldots, r_K \}$ represents the set of reward scores for the $K$ candidate items in the group. The function $\text{mean}(\mathbf{r})$ computes the average reward of the group, and $\text{std}(\mathbf{r})$ calculates the corresponding standard deviation.
Different from conventional GRPO, we use InfoNCE rather than KL divergence as a semantic regularizer. Token-level KL divergence is ill-suited for continuous embedding spaces, whereas InfoNCE directly aligns the learned semantic space with the retrieval objective, leading to more stable and effective optimization for semantic retrieval.

\subsubsection{R-DPO Loss}
\label{sec:supervised-rdpo}
We also introduce a retrieval-specific direct preference optimization objective, termed R-DPO, which adapts the DPO framework for vector retrieval tasks by removing the reference policy constraint. It provides explicit supervision for fine-grained semantic distinctions. 
Specifically, positive pairs $(q_i, d_i)$ are extracted from real user click logs, while negatives $d_{\text{hn}}$ are hard negatives mined by the SID-based approach in Section~\ref{sec:supervised-hn-masking}. Given these pairs, the R-DPO loss is formulated as:
\begin{equation}
\mathcal{L}_{\text{R-DPO}} = -\log \sigma\left( \beta \left( s(q_i, d_i) - s(q_i, d_{\text{hn}}) \right) \right),
\label{eq:rdpo_loss}
\end{equation}
where $s(q, d)$ is the relevance scoring function, $\sigma(\cdot)$ is the sigmoid function, and $\beta$ is a temperature parameter controlling the margin strength.
Unlike conventional DPO that regularizes against a reference policy, our R-DPO focuses purely on maximizing the scoring margin between positive and hard negative examples. This design is particularly suitable for retrieval tasks where we aim to explicitly widen the relevance gap between semantically similar items.

\subsubsection{SSR-GRPO: Joint Optimization Framework}
\label{sec:ssr-grpo-joint}
By combining the loss formulas of SID-R-GRPO and R-DPO, we obtain the final loss formulation of SSR-GRPO. To mitigate gradient conflicts and eliminate manual tuning of loss weights, we use uncertainty-based dynamic weighting \cite{kendall2018multi} with trainable scalars $\sigma_1, \sigma_2 > 0$:

\begin{equation}
\mathcal{L}_{\text{SSR-GRPO}} = \frac{1}{2\sigma_1^2} \mathcal{L}_{\text{SID-R-GRPO}} + \frac{1}{2\sigma_2^2} \mathcal{L}_{\text{R-DPO}} + \log(\sigma_1 \sigma_2),
\label{eq:ssrgrpo_dpo_loss}
\end{equation}
Here, $\frac{1}{2\sigma_i^2}$ acts as an adaptive weight: tasks with higher noise or difficulty (larger $\sigma$) are down-weighted to prevent gradient dominance, while $\log(\sigma)$ serves as a regularizer. This mechanism ensures stable convergence by dynamically balancing the high-variance RL signals from SID-R-GRPO with the deterministic supervised gradients from R-DPO, effectively aligning their optimization trajectories without hyperparameter sensitivity.

Based on the above descriptions and optimization steps, the overall pipeline of SSR-GRPO can be summarized and organized in Figure \ref{fig:ssr} and
the specific implementation steps are as follows:
\begin{enumerate}
  \item \textbf{Candidate Sample Selection:} Retrieve top-$k$ candidates via nearest-neighbor search over query and item representations.
  \item \textbf{Dense-Sparse Perspective Reward Computation:} The multi-objective rewards are computed by Equation~\eqref{eq:reward_components}. The relevance reward is comprehensively derived from the dual-perspective dense-sparse relevance reward scores introduced in ~\ref{subsec:sid-scoring}.

  \item \textbf{R-DPO data construction:} \label{step:bpr} 
 Construct R-DPO pairs using hard negative samples obtained from the hierarchical SID-based approach and a related-but-unexposed-item strategy introduced in ~\ref{sec:supervised-rdpo}.

  \item \textbf{Masking Mechanism Calculation} 
  Relevance-based hard negatives are employed as supervision to calculate a masking function by Equation ~\eqref{eq:ssr_mask_definition} for filtering out easy negatives in the group.
  \item \textbf{SSR-GRPO Loss Optimization} 
    Minimize Eq.~\eqref{eq:ssrgrpo_dpo_loss}, jointly updating model parameters $\theta$ and uncertainty scalars $\sigma_{1,2}$.
\end{enumerate}

\section{Experiments}
\label{sec:exp}

In this section, we introduce our industrial-scale dataset from a real-world scenario, evaluation metrics, experimental configurations, and model hyperparameters. We present the results of our proposed model through both offline evaluations and online A/B tests.

\begin{table*}[t]
\centering
\caption{Overall performance comparison on the different methods in terms of Hitrate@4k and Goodrate@100}
\label{tab:main_results}
\begin{tabular}{lcccc}
\hline
\multirow{2}{*}{\textbf{Model}} & \multicolumn{2}{c}{\textbf{HR@4k}} & \multicolumn{2}{c}{\textbf{GR@100}} \\
 & \textbf{General} & \textbf{Long-Tail} & \textbf{General} & \textbf{Long-Tail} \\
\hline
StructBERT & 0.7166 & 0.4019 & 0.8837 & 0.5732 \\
Qwen2.5-1.5B &  0.7466 & 0.4189 & 0.8894 & 0.6415 \\
Qwen3-0.6B & 0.7495 & 0.4330 & 0.8958 & 0.6524 \\
Tbstars-3B & 0.7577 & 0.4506 & 0.9088 & 0.6583 \\
\hline
Tbstars-3B SFT (base) & 0.7872 & 0.4754 & 0.9174 & 0.6941 \\
Tbstars-3B SFT + R-DPO & 0.7916 & 0.5857 & 0.9261 & 0.7014 \\
Tbstars-3B SFT + R-GRPO & 0.8158 & 0.5857 & 0.9305 & 0.7126 \\
Tbstars-3B SFT + SSR-GRPO & \textbf{0.8218} & \textbf{0.5943} & \textbf{0.9372} & \textbf{0.7195} \\

Tbstars-3B R-GRPO only & 0.7616 & 0.4492 & 0.9213 & 0.6820 \\
Tbstars-3B SSR-GRPO only & 0.7628 & 0.4562 & 0.9230 & 0.6844 \\
\hline
\end{tabular}
\end{table*} 

\subsection{Experimental Setup}
\label{subsec:setup}

\subsubsection{Dataset and Metrics}
We utilized over 0.3 billion Tmall click-and-purchase logs as positive samples. The model undergoes two-stage training: SFT with InfoNCE loss on the full three-month dataset, followed by SSR-GRPO optimization on recent 15-day interactions. Performance is evaluated via Hitrate@4K (HR@4k) and Goodrate@100 (GR@100)~\cite{li2021embedding,zhang2020towards}.

\begin{itemize}
\item \textbf{Hitrate@4k}: This metric measures the recall rate of the ground-truth items within the top-4000 candidates retrieved by the dense retrieval model. It evaluates the system's ability to successfully recover all relevant items.
 \item \textbf{Goodrate@100}: This metric is defined as the precision of items rated as "Good" (i.e., score 1 for "Related") among the top-100 retrieved results. It quantifies the relevance quality of the top results presented to users.
\end{itemize}
Next, we provide a concise description of the two offline test sets:
\begin{itemize}
\item \textbf{General Retrieval Test Set}: To avoid overfitting, this set is built from a distinct e-commerce scenario's user interactions. It contains clicked and purchased items, filtered by relevance model and human annotation, with over 450,000 queries and 1,330,000 query-item pairs. 
\item \textbf{Long-Tail Query Test Set}: Constructed by collecting long-tail queries from end-to-end search logs across other e-commerce scenarios. These queries, which can be viewed as out-of-distribution data, were annotated using both relevance models and manual labeling, followed by rigorous filtering. The final test set contains over 30,000 queries and 90,000 query-item pairs.

\end{itemize}

\subsubsection{Implementation and Deployment Details}
For the SFT stage, we use Tbstars-3B (an internal 3B-parameter model) as our backbone encoder. Training employs 32 GPUs with 96GB memory each, batch size 140, AdamW optimizer (lr=1e-5), max input length 100 tokens, and 128-dimensional L2-normalized embeddings for 1 epoch. In the SSR-GRPO phase, we use 15 days of data with reduced learning rate 5e-6, maintaining other hyperparameters from SFT. The InfoNCE loss weight $\beta$ in Equation~\eqref{sidrgrpo_loss} is set to 0.5 for best performance, and training runs for 3 epochs.

\textbf{Offline Training Process:} During offline training, we input item titles and basic attribute texts to the LLM for the item side, and search queries to the LLM for the query side, following the SSR-GRPO training procedure.

\textbf{Online Deployment:} The trained SSR-GRPO LLM is used to infer embeddings for the entire product pool, which are cached online. The LLM is then deployed to production, where it performs real-time inference on user search queries to generate query vectors. These vectors retrieve Top-K items from the cached product embeddings, which are subsequently passed to downstream modules.

\subsubsection{Baseline Models}
\label{subsec:baselines}

We compare SSR-GRPO against three categories of baselines: the production Tmall baseline (Tbstars-3B SFT), the primary competitor R-GRPO, and several state-of-the-art text embedding models.
The specific baselines are as follows:
\begin{itemize}
    \item StructBERT \cite{wang2019structbert}: A structure-aware encoder that integrates linguistic structural priors (e.g., word order and sentence coherence) into pre-training, with 0.1 billion parameters.
    
    \item Qwen3-0.6B: The Qwen3-Embedding\-0.6B model \cite{zhang2025qwen3} is designed for text embedding and ranking tasks.
    
    \item Qwen2.5-1.5B \cite{team2024qwen2}: One of the Qwen2.5 series with 1.5 billion parameters, designed for generating long texts and understanding structured data.

    \item Tbstars-3B: Dense retrieval model trained based on an internal closed-source Tbstars with 3B parameters, modifying to bi-directional attention.
    
    \item Tbstars-3B SFT (base): This model is built upon Tbstars-3B and fine-tuned on Tmall search logs using the InfoNCE loss, which can be viewed as the first phase of SSR-GRPO in Section~\ref{sec:infonce-loss}. It had been deployed in the online system, serving hundreds of millions of users. 

    \item Tbstars-3B SFT + R-GRPO: The R-GRPO training framework which includes SFT and R-GRPO, representing a direct ablation of our SSR-GRPO framework.
\end{itemize}

\subsection{Offline Evaluation}
\label{sec:exp results}

The experimental results are shown in Table~\ref{tab:main_results}. The proposed Tbstars-3B SFT + SSR-GRPO achieves the best performance across all evaluation settings, obtaining the highest scores on both General Retrieval Test and Long-Tail Query test sets. This demonstrates the effectiveness of our SSR-GRPO framework in enhancing both retrieval coverage and relevance quality. Among the baseline models, Tbstars-3B outperforms both Qwen2.5 and Qwen3-0.6B across all metrics, validating its superior architecture for e-commerce retrieval tasks. The Tbstars-3B SFT (base) model shows significant improvements over the pre-trained Tbstars-3B, confirming the value of domain-specific fine-tuning on Tmall search logs. The integration of reinforcement learning methods consistently improves upon the SFT baseline. Tbstars-3B SFT + R-GRPO shows substantial gains, while our proposed SSR-GRPO demonstrates clear advantages over R-GRPO in all metrics (especially showing a 0.6\% gain in HR@4k and 0.86\% in Long-Tail HR@4k), validating the effectiveness of incorporating supervised signals and SIDs for enhancing relevance reward. Notably, the standalone GRPO variants without SFT initialization show limited performance, indicating the SFT stage remains essential for effective training.

\begin{table}[ht]
\centering
\caption{Ablation study on the contributions of SSR-GRPO components.}
\label{tab:ablation_components}
\begin{tabular}{lcc}
\toprule
\textbf{Model} & \textbf{Hitrate@4k} & \textbf{GR@100} \\
\midrule
SFT+SSR-GRPO (Full) & +0.00\% & +0.00\% \\
\hline
(w/o Dual-Perspective Reward) & -0.17\% & -0.25\% \\
(w/o R-DPO Loss) & -0.49\% & -0.36\% \\
(w/o Masking Function) & -0.12\% & -0.08\% \\
(w/o Dynamic Weights) & -0.15\% & -0.17\%  \\
\bottomrule
\end{tabular}
\end{table}

\subsection{Ablation Study}
\label{sec:ablation study}
In this section, we only present HR@4k on the general evaluation set and GR@100 on the long-tail dataset to provide a more intuitive experimental comparison.

\subsubsection{Analysis of SSR-GRPO Components}

To validate the effectiveness of each component in SSR-GRPO compared to R-GRPO, we conducted ablation experiments on SSR-GRPO. To perform the ablation studies, we removed each of the three components one at a time. Dual-Perspective Reward represents the component described in Section~\ref{subsec:sid-scoring}, R-DPO loss means Equation~\eqref{eq:rdpo_loss}, Masking Function is Equation~\eqref{eq:ssr_mask_definition} and Dynamic Weights means the uncertainty-based dynamic weighting in ~\eqref{eq:ssrgrpo_dpo_loss}. As Table ~\ref{tab:ablation_components} shows, each ablated configuration resulted in a performance degradation across evaluation metrics. Most notably, the removal of the R-DPO Loss caused the most significant drop in HR@4k by 0.49\%, underscoring the critical and irreplaceable importance of hard negative mining. This finding also demonstrates the complementary effect of R-DPO and R-GRPO in the task. In contrast to methods solely dependent on reward computation, which are plagued by high variance, the introduction of R-DPO furnishes the training process with stable and well-defined gradients, thereby effectively constraining the overall optimization trajectory. 

Additionally, SID integration contributes consistently to both metrics, particularly benefiting relevance judgment as evidenced by the 0.25\% drop in GR@100 when removed. This highlights the value of combining dense and sparse representations for a comprehensive semantic understanding. Although the masking mechanism contributes the least among the four components, the 0.12\% decrease in Hitrate@4K nonetheless validates its effectiveness. Moreover, this masking approach efficiently filters out unnecessary simple negative samples, thereby improving computational efficiency during training.

\begin{table}[ht]
\centering
\caption{Ablation results on Relevance Reward}
\label{tab:ablation_relevance}
\begin{tabular}{lcc}
\toprule
\textbf{Model} & \textbf{HR@4k} & \textbf{GR@100} \\
\midrule
R-GRPO (w TaoSR1) & 0.8152 & 0.7136 \\
R-GRPO (w Dense) & 0.8093 & 0.7108 \\
R-GRPO (w Sparse SID) & 0.8154 & 0.7143 \\
R-GRPO (w Dual-Perspective SID) & \textbf{0.8161} & \textbf{0.7165} \\
\bottomrule
\end{tabular}
\end{table}

\subsubsection{Effect of Relevance Reward}

SSR-GRPO replaces the computationally heavy TaoSR1 LLM in R-GRPO with a lightweight dual-perspective reward mechanism. As shown in Table~\ref{tab:ablation_relevance}, the Sparse SID-only variant achieves performance comparable to, and even slightly better than TaoSR1 (HR@4k: 0.8154 vs. 0.8152). This demonstrates that representing items as discrete semantic clusters via SIDs provides a robust and unbiased signal for GRPO reward estimation, effectively capturing coarse-grained relevance without the bias inherent in LLM-based judges. While the Dense-only variant shows limited capability, the Dual-Perspective fusion further boosts performance by integrating fine-grained dense nuances. In our implementation, we set the balancing coefficient $\alpha=0.8$ in Eq.~\eqref{eq:relevance_fusion}, which yields the optimal trade-off between dense semantic smoothness and sparse hierarchical discrimination.

\subsubsection{Case Study}
\label{subsec:case_study}

\begin{figure}[t]
    \centering
    \includegraphics[width=0.95\linewidth]{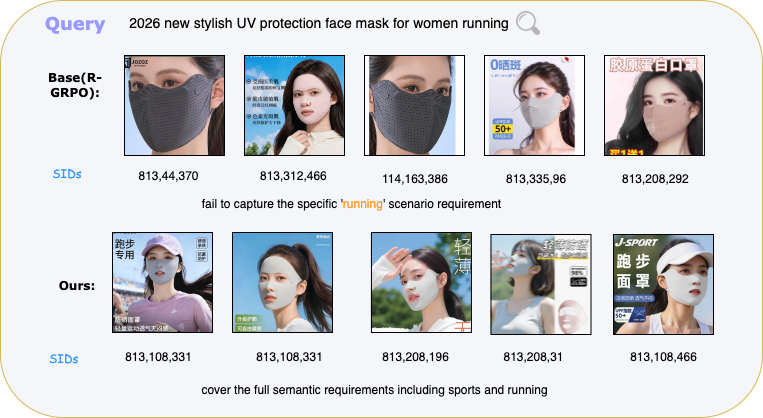} 
    \caption{Comparison of retrieval results: Top-exposed items from online A/B tests for our model and R-GRPO.}
    \label{fig:case_study}
\end{figure}

To qualitatively assess the semantic alignment capability of SSR-GRPO, we analyze a complex multi-attribute query: "2026 new stylish UV protection face mask for women running". As illustrated in Figure~\ref{fig:case_study}, the baseline R-GRPO retrieves items that are visually appealing but fail to capture the specific "running" scenario. This indicates that R-GRPO is prone to Reward Hacking when the Top-K candidate pool lacks perfect matches. In contrast, SSR-GRPO comprehensively covers the full semantic requirements, retrieving masks that balance high aesthetics with running-specific functionalities (e.g., breathability and secure fit).

The superior performance on this case is attributed to the SID-guided optimization mechanism. By leveraging hierarchical SIDs to construct hard negatives, the R-DPO component forces the model to learn fine-grained semantic distinctions, preventing collapse into coarse-grained local optima. Meanwhile, the masking function filters out trivial negatives. A compelling evidence of this improved coherence is the distribution of SIDs in the retrieved results. As shown in Figure~\ref{fig:case_study}, the baseline SIDs are highly dispersed across different semantic clusters (e.g., the second index level varies significantly: $335, 208, 163, \dots$). Conversely, SSR-GRPO results exhibit strong clustering consistency, with the second index level concentrating on only two values ($208, 108$). This concentration confirms our method effectively reduces semantic noise and aligns the retrieved items with the user's precise multi-faceted intent.

\subsection{Online Evaluation}
\label{subsec:online_results}

We deployed the semantic retrieval model in production search on
TmallAPP, a major mobile e-commerce application. Our online evaluation focused on key business metrics: UCTCVR, GMV, Ex-Imp(the proportion of items seen by users that are solely retrieved by the semantic deep retrieval channel in our multi-channel framework) and Goodrate (the mean relevance ratio derived by sampling search queries and corresponding items from search logs).

\begin{table}[ht]
\centering
\caption{Online A/B testing results (relative improvement over baseline).}
\label{tab:online_results}
\begin{tabular}{lcccc}
\toprule
\textbf{Model} & \textbf{UCTCVR} & \textbf{GMV} & \textbf{Ex. Imp.} & \textbf{Goodrate} \\
\midrule
Base & - & - & - & - \\
R-GRPO & +0.38\% & +1.01\% & +0.37\% & +0.20\% \\
SSR-GRPO & \textbf{+0.99\%} & \textbf{+1.40\%} & \textbf{+0.48\%} & \textbf{+0.61\%} \\
\bottomrule
\end{tabular}
\end{table}
The online A/B test was conducted over a period of two weeks to ensure statistical reliability. Compared to the strong baseline R-GRPO, SSR-GRPO achieved statistically significant improvements, with a 0.61\% lift in UCTCVR and a 0.39\% increase in GMV. These results further validate the business value of our proposed method in a real-world production environment.

\section{Conclusion}
\label{sec:conclude}

This paper presents SSR-GRPO, a novel RL framework for dense
retrieval that leverages hierarchical SIDs to provide unbiased rel-
evance rewards, replacing costly LLMs. By integrating SID-based
hard negative mining for R-DPO supervision and a masking mech-
anism to filter noise in R-GRPO, our method ensures stable and
precise optimization. Experiments show SSR-GRPO significantly outperforms state-of-the-art baselines, offering superior semantic alignment and robust business value.

\bibliographystyle{ACM-Reference-Format}
\bibliography{myrefs}

@inproceedings{sondhi2018taxonomy,
  title={A taxonomy of queries for e-commerce search},
  author={Sondhi, Parikshit and Sharma, Mohit and Kolari, Pranam and Zhai, ChengXiang},
  booktitle={The 41st International ACM SIGIR Conference on Research \& Development in Information Retrieval},
  pages={1245--1248},
  year={2018}
}

@inproceedings{huang2020embedding,
  title={Embedding-based retrieval in facebook search},
  author={Huang, Jui-Ting and Sharma, Ashish and Sun, Shuying and Xia, Li and Zhang, David and Pronin, Philip and Padmanabhan, Janani and Ottaviano, Giuseppe and Yang, Linjun},
  booktitle={Proceedings of the 26th ACM SIGKDD International Conference on Knowledge Discovery \& Data Mining},
  pages={2553--2561},
  year={2020}
}

@inproceedings{huang2013learning,
  title={Learning deep structured semantic models for web search using clickthrough data},
  author={Huang, Po-Sen and He, Xiaodong and Gao, Jianfeng and Deng, Li and Acero, Alex and Heck, Larry},
  booktitle={Proceedings of the 22nd ACM international conference on Information \& Knowledge Management},
  pages={2333--2338},
  year={2013}
}

@inproceedings{devlin2019bert,
  title={Bert: Pre-training of deep bidirectional transformers for language understanding},
  author={Devlin, Jacob and Chang, Ming-Wei and Lee, Kenton and Toutanova, Kristina},
  booktitle={Proceedings of the 2019 conference of the North American chapter of the association for computational linguistics: human language technologies, volume 1 (long and short papers)},
  pages={4171--4186},
  year={2019}
}

@article{liu2019roberta,
  title={Roberta: A robustly optimized bert pretraining approach},
  author={Liu, Yinhan and Ott, Myle and Goyal, Naman and Du, Jingfei and Joshi, Mandar and Chen, Danqi and Levy, Omer and Lewis, Mike and Zettlemoyer, Luke and Stoyanov, Veselin},
  journal={arXiv preprint arXiv:1907.11692},
  year={2019}
}

@article{oord2018representation,
  title={Representation learning with contrastive predictive coding},
  author={Oord, Aaron van den and Li, Yazhe and Vinyals, Oriol},
  journal={arXiv preprint arXiv:1807.03748},
  year={2018}
}

@article{robinson2020contrastive,
  title={Contrastive learning with hard negative samples},
  author={Robinson, Joshua and Chuang, Ching-Yao and Sra, Suvrit and Jegelka, Stefanie},
  journal={arXiv preprint arXiv:2010.04592},
  year={2020}
}

@article{lee2025gemini,
  title={Gemini embedding: Generalizable embeddings from gemini},
  author={Lee, Jinhyuk and Chen, Feiyang and Dua, Sahil and Cer, Daniel and Shanbhogue, Madhuri and Naim, Iftekhar and {\'A}brego, Gustavo Hern{\'a}ndez and Li, Zhe and Chen, Kaifeng and Vera, Henrique Schechter and others},
  journal={arXiv preprint arXiv:2503.07891},
  year={2025}
}

@article{zhang2025qwen3,
  title={Qwen3 Embedding: Advancing Text Embedding and Reranking Through Foundation Models},
  author={Zhang, Yanzhao and Li, Mingxin and Long, Dingkun and Zhang, Xin and Lin, Huan and Yang, Baosong and Xie, Pengjun and Yang, An and Liu, Dayiheng and Lin, Junyang and others},
  journal={arXiv preprint arXiv:2506.05176},
  year={2025}
}

@article{jiang2025deepretrieval,
  title={Deepretrieval: Hacking real search engines and retrievers with large language models via reinforcement learning},
  author={Jiang, Pengcheng and Lin, Jiacheng and Cao, Lang and Tian, Runchu and Kang, SeongKu and Wang, Zifeng and Sun, Jimeng and Han, Jiawei},
  journal={arXiv preprint arXiv:2503.00223},
  year={2025}
}

@article{zeng2025optimizing,
  title={Optimizing Generative Ranking Relevance via Reinforcement Learning in Xiaohongshu Search},
  author={Zeng, Ziyang and Jing, Heming and Chen, Jindong and Li, Xiangli and Liu, Hongyu and He, Yixuan and Li, Zhengyu and Sun, Yige and Xie, Zheyong and Yang, Yuqing and others},
  journal={arXiv preprint arXiv:2512.00968},
  year={2025}
}

@article{shao2024deepseekmath,
  title={Deepseekmath: Pushing the limits of mathematical reasoning in open language models},
  author={Shao, Zhihong and Wang, Peiyi and Zhu, Qihao and Xu, Runxin and Song, Junxiao and Bi, Xiao and Zhang, Haowei and Zhang, Mingchuan and Li, YK and Wu, Yang and others},
  journal={arXiv preprint arXiv:2402.03300},
  year={2024}
}

@article{liu2025taosearchemb,
  title={TaoSearchEmb: A Multi-Objective Reinforcement Learning Framework for Dense Retrieval in Taobao Search},
  author={Liu, Xingxian and Li, Dongshuai and Wen, Tao and Wan, Jiahui and Ling, Gui and Lv, Fuyu and Ou, Dan and Tang, Haihong},
  journal={arXiv preprint arXiv:2511.13885},
  year={2025}
}

@article{liu2025deepseek,
  title={Deepseek-v3. 2: Pushing the frontier of open large language models},
  author={Liu, Aixin and Mei, Aoxue and Lin, Bangcai and Xue, Bing and Wang, Bingxuan and Xu, Bingzheng and Wu, Bochao and Zhang, Bowei and Lin, Chaofan and Dong, Chen and others},
  journal={arXiv preprint arXiv:2512.02556},
  year={2025}
}

@inproceedings{liu2021que2search,
  title={Que2search: fast and accurate query and document understanding for search at facebook},
  author={Liu, Yiqun and Rangadurai, Kaushik and He, Yunzhong and Malreddy, Siddarth and Gui, Xunlong and Liu, Xiaoyi and Borisyuk, Fedor},
  booktitle={Proceedings of the 27th ACM SIGKDD Conference on Knowledge Discovery \& Data Mining},
  pages={3376--3384},
  year={2021}
}

@inproceedings{li2021embedding,
  title={Embedding-based product retrieval in taobao search},
  author={Li, Sen and Lv, Fuyu and Jin, Taiwei and Lin, Guli and Yang, Keping and Zeng, Xiaoyi and Wu, Xiao-Ming and Ma, Qianli},
  booktitle={Proceedings of the 27th ACM SIGKDD Conference on Knowledge Discovery \& Data Mining},
  pages={3181--3189},
  year={2021}
}

@inproceedings{guo2020detext,
  title={Detext: A deep text ranking framework with bert},
  author={Guo, Weiwei and Liu, Xiaowei and Wang, Sida and Gao, Huiji and Sankar, Ananth and Yang, Zimeng and Guo, Qi and Zhang, Liang and Long, Bo and Chen, Bee-Chung and others},
  booktitle={Proceedings of the 29th ACM international conference on information \& knowledge management},
  pages={2509--2516},
  year={2020}
}

@inproceedings{niu2020dual,
  title={A dual heterogeneous graph attention network to improve long-tail performance for shop search in e-commerce},
  author={Niu, Xichuan and Li, Bofang and Li, Chenliang and Xiao, Rong and Sun, Haochuan and Deng, Hongbo and Chen, Zhenzhong},
  booktitle={Proceedings of the 26th ACM SIGKDD International Conference on Knowledge Discovery \& Data Mining},
  pages={3405--3415},
  year={2020}
}

@article{gao2021simcse,
  title={Simcse: Simple contrastive learning of sentence embeddings},
  author={Gao, Tianyu and Yao, Xingcheng and Chen, Danqi},
  journal={arXiv preprint arXiv:2104.08821},
  year={2021}
}

@inproceedings{ma2024fine,
  title={Fine-tuning llama for multi-stage text retrieval},
  author={Ma, Xueguang and Wang, Liang and Yang, Nan and Wei, Furu and Lin, Jimmy},
  booktitle={Proceedings of the 47th International ACM SIGIR Conference on Research and Development in Information Retrieval},
  pages={2421--2425},
  year={2024}
}

@article{lee2024nv,
  title={Nv-embed: Improved techniques for training llms as generalist embedding models},
  author={Lee, Chankyu and Roy, Rajarshi and Xu, Mengyao and Raiman, Jonathan and Shoeybi, Mohammad and Catanzaro, Bryan and Ping, Wei},
  journal={arXiv preprint arXiv:2405.17428},
  year={2024}
}

@article{schulman2017proximal,
  title={Proximal policy optimization algorithms},
  author={Schulman, John and Wolski, Filip and Dhariwal, Prafulla and Radford, Alec and Klimov, Oleg},
  journal={arXiv preprint arXiv:1707.06347},
  year={2017}
}

@article{rafailov2023direct,
  title={Direct preference optimization: Your language model is secretly a reward model},
  author={Rafailov, Rafael and Sharma, Archit and Mitchell, Eric and Manning, Christopher D and Ermon, Stefano and Finn, Chelsea},
  journal={Advances in neural information processing systems},
  volume={36},
  pages={53728--53741},
  year={2023}
}

@article{rajput2023recommender,
  title={Recommender systems with generative retrieval},
  author={Rajput, Shashank and Mehta, Nikhil and Singh, Anima and Hulikal Keshavan, Raghunandan and Vu, Trung and Heldt, Lukasz and Hong, Lichan and Tay, Yi and Tran, Vinh and Samost, Jonah and others},
  journal={Advances in Neural Information Processing Systems},
  volume={36},
  pages={10299--10315},
  year={2023}
}

@article{wu2024hi,
  title={Hi-gen: Generative retrieval for large-scale personalized e-commerce search},
  author={Wu, Yanjing and Feng, Yinfu and Wang, Jian and Zhou, Wenji and Ye, Yunan and Xiao, Rong and Xiao, Jun},
  journal={arXiv preprint arXiv:2404.15675},
  year={2024}
}

@inproceedings{liu2025generative,
  title={Generative recommender with end-to-end learnable item tokenization},
  author={Liu, Enze and Zheng, Bowen and Ling, Cheng and Hu, Lantao and Li, Han and Zhao, Wayne Xin},
  booktitle={Proceedings of the 48th International ACM SIGIR Conference on Research and Development in Information Retrieval},
  pages={729--739},
  year={2025}
}

@inproceedings{lin2025order,
  title={Order-agnostic identifier for large language model-based generative recommendation},
  author={Lin, Xinyu and Shi, Haihan and Wang, Wenjie and Feng, Fuli and Wang, Qifan and Ng, See-Kiong and Chua, Tat-Seng},
  booktitle={Proceedings of the 48th international ACM SIGIR conference on research and development in information retrieval},
  pages={1923--1933},
  year={2025}
}

@inproceedings{esser2021taming,
  title={Taming transformers for high-resolution image synthesis},
  author={Esser, Patrick and Rombach, Robin and Ommer, Bjorn},
  booktitle={Proceedings of the IEEE/CVF conference on computer vision and pattern recognition},
  pages={12873--12883},
  year={2021}
}

@inproceedings{lee2022autoregressive,
  title={Autoregressive image generation using residual quantization},
  author={Lee, Doyup and Kim, Chiheon and Kim, Saehoon and Cho, Minsu and Han, Wook-Shin},
  booktitle={Proceedings of the IEEE/CVF conference on computer vision and pattern recognition},
  pages={11523--11532},
  year={2022}
}

@inproceedings{zhang2020towards,
  title={Towards personalized and semantic retrieval: An end-to-end solution for e-commerce search via embedding learning},
  author={Zhang, Han and Wang, Songlin and Zhang, Kang and Tang, Zhiling and Jiang, Yunjiang and Xiao, Yun and Yan, Weipeng and Yang, Wen-Yun},
  booktitle={Proceedings of the 43rd International ACM SIGIR Conference on Research and Development in Information Retrieval},
  pages={2407--2416},
  year={2020}
}

@inproceedings{kong2022multi,
  title={Multi-aspect dense retrieval},
  author={Kong, Weize and Khadanga, Swaraj and Li, Cheng and Gupta, Shaleen Kumar and Zhang, Mingyang and Xu, Wensong and Bendersky, Michael},
  booktitle={Proceedings of the 28th ACM SIGKDD Conference on Knowledge Discovery and Data Mining},
  pages={3178--3186},
  year={2022}
}

@inproceedings{wang2023learning,
  title={Learning multi-stage multi-grained semantic embeddings for e-commerce search},
  author={Wang, Binbin and Li, Mingming and Zeng, Zhixiong and Zhuo, Jingwei and Wang, Songlin and Xu, Sulong and Long, Bo and Yan, Weipeng},
  booktitle={Companion Proceedings of the ACM Web Conference 2023},
  pages={411--415},
  year={2023}
}

@article{dong2025taosr1,
  title={TaoSR1: The thinking model for e-commerce relevance search},
  author={Dong, Chenhe and Yao, Shaowei and Jiao, Pengkun and Yang, Jianhui and Jin, Yiming and Huang, Zerui and Zhou, Xiaojiang and Ou, Dan and Tang, Haihong and Zheng, Bo},
  journal={arXiv preprint arXiv:2508.12365},
  year={2025}
}

@inproceedings{li2023learning,
  title={Learning query-aware embedding index for improving e-commerce dense retrieval},
  author={Li, Mingming and Yuan, Chunyuan and Wang, Binbin and Zhuo, Jingwei and Wang, Songlin and Liu, Lin and Xu, Sulong},
  booktitle={Proceedings of the 46th International ACM SIGIR Conference on Research and Development in Information Retrieval},
  pages={3265--3269},
  year={2023}
}

@article{wang2019structbert,
  title={Structbert: Incorporating language structures into pre-training for deep language understanding. arXiv},
  author={Wang, W and Bi, B and Yan, M and Wu, C and Bao, Z and Xia, J and Peng, L and Si, L},
  journal={arXiv preprint arXiv:1908.04577},
  year={2019}
}

@article{team2024qwen2,
  title={Qwen2 technical report},
  author={Team, Qwen and others},
  journal={arXiv preprint arXiv:2407.10671},
  volume={2},
  number={3},
  year={2024}
}

@article{wu2024llm,
  title={Llm-augmented retrieval: Enhancing retrieval models through language models and doc-level embedding},
  author={Wu, Mingrui and Cao, Sheng},
  journal={arXiv preprint arXiv:2404.05825},
  year={2024}
}

@inproceedings{kendall2018multi,
  title={Multi-task learning using uncertainty to weigh losses for scene geometry and semantics},
  author={Kendall, Alex and Gal, Yarin and Cipolla, Roberto},
  booktitle={Proceedings of the IEEE conference on computer vision and pattern recognition},
  pages={7482--7491},
  year={2018}
}

\section*{GenAI Usage Disclosure}
During the preparation of this work, GenAI tools were used solely
to improve the spelling and grammar of the author-written text. All ideas, theoretical frameworks, experimental designs, results, and conclusions are the original work of the authors.

\end{document}